\def\arxivversion{1}
\documentclass[conference]{IEEEtran}
\IEEEoverridecommandlockouts

\usepackage{amsmath,amssymb,amsfonts}
\usepackage{graphicx}
\usepackage{multirow}
\usepackage{enumitem}
\usepackage{cite}

\ifdefined\arxivversion
  \newcommand{\arxivnotice}{%
    © 2025 IEEE. Personal use of this material is permitted. Permission from
    IEEE must be obtained for all other uses, in any current or future media,
    including reprinting/republishing this material for advertising or
    promotional purposes, creating new collective works, for resale or
    redistribution to servers or lists, or reuse of any copyrighted component
    of this work in other works.\newline
    This is the accepted author manuscript. Published in the \emph{2025 7th
    IEEE International Conference on Artificial Intelligence Circuits and
    Systems (AICAS)}, 2025, doi: 10.1109/AICAS64808.2025.11173151.}
\fi

\begin{document}

%%
%% The "title" command has an optional parameter,
%% allowing the author to define a "short title" to be used in page headers.
\title{Non-uniform Memory Partitioning For Low-Power Spiking Neural Networks}

%%
%% The "author" command and its associated commands are used to define
%% the authors and their affiliations.
%% Of note is the shared affiliation of the first two authors, and the
%% "authornote" and "authornotemark" commands
%% used to denote shared contribution to the research.

\author{
Simon Richter, Darío Fernández Khatiboun, Maryam Sadeghi, Milad Zamani, Farshad Moradi
\\{Department of Electrical and Computer Engineering, Aarhus University, Denmark}
%\\{Email: \{siri, dariofk, sadeghi, mzamani, moradi\}@ece.au.dk}
\ifdefined\arxivversion
  \thanks{\arxivnotice}
\fi
}

\graphicspath{{figures/}}

\maketitle

%%
%% The abstract is a short summary of the work to be presented in the
%% article.
\begin{abstract}

%Spiking Neural Networks (SNNs) naturally excel in processing temporally rich and sparse data. However, because of their time-stepped processing, memory access, specifically to synaptic weights stored in SRAM (Static Random-Access Memory), tends to dominate total power consumption. To address this issue, without incurring a large area overhead, we propose to leverage the varying average firing activity of neurons in the network to efficiently allocate synaptic weights to a on-chip memory consisting multiple non-uniformly sized memory banks. By assigning synaptic weights of frequently firing neurons to more shallow memories that are less expensive to access, the average power consumption of the memory is decreased. A automatic exploration of different memory arrangements based on application requirements and hardware constraints is performed to find optimal memory configuration. We show that our architecture can achieve a synaptic weight memory access power reduction of up to 61\% compared to conventional single synaptic weight memory, with a 2.1$\times$ lower area overhead, as compared to a traditional uniformly partitioned memory that achieves a comparable reduction.

Spiking Neural Networks (SNNs) naturally excel in processing temporally rich and sparse data. However, because of their time-stepped processing, memory access, specifically to synaptic weights stored in SRAM (static random-access memory), tends to dominate total power consumption. To address this issue, without incurring a large area overhead, we propose to leverage the greatly varying average firing rate of neurons in the network to efficiently allocate synaptic weights to an on-chip memory consisting of multiple non-uniformly sized memory banks. By assigning weights of frequently firing neurons to shallow, low-access cost memory and less actively accessed weights to deeper, high-density memories, the average power consumption of the synaptic weight memory is decreased without incurring a large area overhead. To benchmark our proposed architecture and find optimal configurations of memory arrangements, we perform an automatic exploration based on application requirements and hardware constraints. For memory designs synthesized in 28-nm CMOS technology, we show that our architecture can achieve a synaptic weight memory access power reduction of up to 61\% compared to a conventional design, with a 2.1$\times$ lower area overhead, as compared to a traditional uniformly partitioned memory bank that achieves a comparable reduction.
\end{abstract} % Should include some results on the increased flexibility with such an approach. I.e. more selection for exact area/power tradeoff. and compare to single uniform SRAM bank. 


\section{Introduction}

Spiking neural networks (SNNs) have emerged as a promising alternative to conventional artificial neural networks (ANNs). SNNs mimic biological neurons by representing information as single-bit spike streams across time, potentially achieving higher energy efficiency compared to their ANN counterparts because of their event-driven data processing. In recent digital SNN implementations, memory access to synaptic weights stored in static random-access memory (SRAM) has been the main contributor to power consumption \cite{ESSCIRC_power_breakdown, jssc-with-power-breakdown, tcas-with-power-breakdown}. To mitigate this issue, weight memory can be subdivided into multiple smaller, uniformly sized, and individually addressable memory banks \cite{ESSCIRC_power_breakdown}, \cite{Benini_memory_design}. However, this imposes overheads, as when the memory is partitioned into increasingly finer segments, the area and power overhead incurred by the SRAM peripheral logic begins to dominate.

\begin{figure}[h]
\centering
\includegraphics[width=\columnwidth]{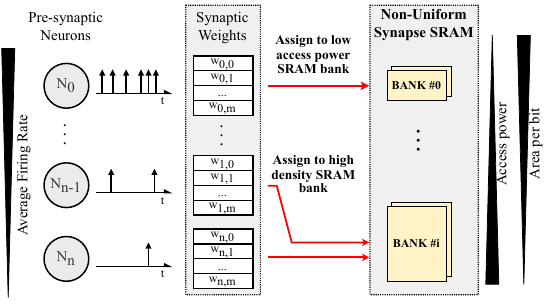}
\caption{Overview of the proposed non-uniform synapse memory architecture with firing-activity based bank allocation.}
\label{fig:intro concept}
\end{figure}

To address this challenge, we propose a non-uniform SRAM partitioning scheme, where the SRAM is divided into banks of varying depths, and weights are allocated to banks based on the firing activity of pre-synaptic neurons, as illustrated in Fig. \ref{fig:intro concept}. In SNNs, due to the nonuniform firing rate of neurons, the frequency at which different synaptic weights are accessed can vary greatly. By storing the weights associated with more frequently firing pre-synaptic neurons in a more shallow, less expensive to access SRAM, and less frequently accessed memories in deeper and denser SRAM, the average memory access cost decreases without a large area overhead. This approach necessitates additional resources to locate the weights in the SRAM banks, which can result in extra latency and logic complexity. To tackle this issue, we propose a neuron sorting technique, where the order in which neurons are iterated over in the SNN controller and address allocation of their weights depends on their expected firing activity. 

\vspace{11pt}
The main contributions in this work can be summarized as: 
\begin{itemize}
  %\item We propose a non-uniform weight memory architecture for synaptic weights based on the observed firing distributions and access pattern, along with an efficient address mapping and sorting scheme to simplify the memory access order.
  \item We propose a non-uniform synaptic weight memory that combines the advantages of both shallow, low-access cost memories with deeper, high-density memories by leveraging the varying fire rate across neurons in SNNs. 
  \item An efficient address mapping and neuron sorting scheme is proposed to simplify the memory access order to individual memory banks.
  \item We investigate the firing rate distribution of neurons across networks for neuromorphic applications and highlight the highly non-uniform distribution of spike firings. 
  \item We perform an automatic exploration of how the memories can be optimally arranged in the proposed non-uniform architecture to achieve maximal access power reduction at the smallest manageable area overhead. Our results show that the proposed memory architecture can reduce access power by upward of 61\%, at a 2.1$\times$ lower area overhead compared to a conventional design.
\end{itemize}

%The proceeding section briefly introduces the issue of memory access dominating power in SNNs. Section \ref{sec:method} describes the proposed architecture with details on the control logic and memory allocation. Section \ref{sec:results} describes the evaluation process and presents the effectiveness of the proposed architecture. Lastly section \ref{sec:discussion} provides a discussion around the results and findings of the design space exploration.

\section{Neuromorphic Accelerators}
\begin{figure}[h]
  \centering
\includegraphics[width=0.95\columnwidth]{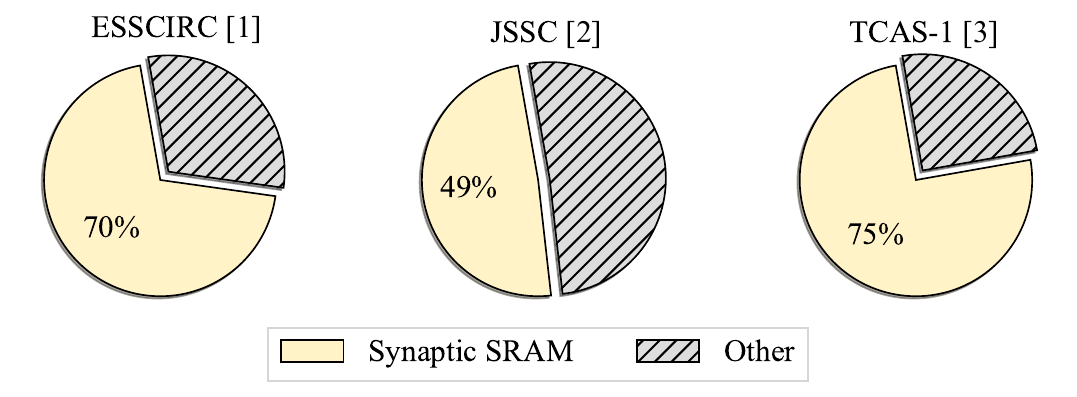}
  \caption{Power breakdown of neuromorphic accelerators.}
  \label{fig:syn power breakdown}
\end{figure}

Various SNN application-specific integrated circuits (ASICs) have been proposed, ranging from analog \cite{imc_snn, dynapse, ronchini2023net} to fully-digital architectures \cite{ESSCIRC_power_breakdown, ODIN, ANP-I}. However, digital designs have shown greater flexibility and scalability. To achieve high densities of neurons and synapses, a neuron processing unit (NPU) is commonly shared, emulating several neurons and their accumulation of synaptic weights and the generation of output spikes \cite{Charlotte_review}. Computing in a time-stepped manner involves updating all neurons within a predefined time interval (timestep), during which synaptic weights are accessed sequentially. Only synaptic weights corresponding to pre-synaptic neurons that have fired in the previous time step need to be accessed. This approach skips unnecessary access to the synapse SRAM and leads to a memory access pattern that depends on the firing rate of neurons in the network. Fig. \ref{fig:syn power breakdown} illustrates the power breakdown of existing digital SNN accelerator designs, highlighting that access to synaptic SRAM dominates the total power consumption of the system. 

\section{Proposed Memory Architecture}
\label{sec:method}
\subsection{Memory Organization and Address Allocation}
\label{sec:address allocation}
We propose partitioning the synaptic weight memory into multiple, individually addressable memory banks. Each bank consists of a single foundry-provided and pre-compiled SRAM macro with peripheral logic. From the SNN controller's perspective, the non-uniform synapse SRAM is perceived as a single, continuous, address space with the combined depth of all concatenated banks. The banks are organized in such a way that the lowest ranges of addresses are formed by and map to the shallowest memory banks with the lowest access power. We assume a generic controller architecture similar to existing SNN designs, where first the membrane potential of a given neuron is loaded from a Neuron SRAM, and then all weights mapping to (firing) spikes in the current neuron receptive field are loaded and processed. By storing more frequently accessed weights in more shallow memory banks, the average power consumption from memory access can be reduced. 

To avoid the requirement for look-up logic for the controller to determine in which bank of the non-uniform memory a given synaptic weight is stored, which would incur extra complexity and latency, we propose a firing activity-based sorting and address allocation scheme of neurons and synaptic weights. After training, the network is profiled in terms of memory accesses based on firing rates, the network topology, and the accelerator architecture for address allocation. 

The implementation of this scheme for of fully connected layer is straightforward, as all post-synaptic neurons are connected to any given pre-synaptic neuron. All synaptic weights originating from the most actively firing neurons are assigned to the shallowest banks of the non-uniform memory. With the pre-synaptic neuron ID forming the least significant bits (LSBs) of the weight address, weights can be allocated into memory in a way that follows the recorded firing activity of the pre-synaptic neurons. This approach ensures that when the address increases, the expected average access rate to that address decreases, mapping to larger banks in the non-uniform memory.

For convolutional layers, more care has to be taken during the assignment of synaptic weights to addresses in the non-uniform memory, as multiple neurons share the same kernel weights. We assume a controller architecture similar to that used for the fully connected layers. For the convolutional layers, the receptive field of a given neuron depends on the position of the post-synaptic neuron in the output feature ($x_{out}$, $y_{out}$, $ch_{out}$). Since the same kernels are applied to the entire input region ($x_{in},  y_{in}$), the access rate of different weights within the same kernel will be roughly the same (ignoring cases at the edge of the image where only part of the kernel may be applied to the input region). The same applies to kernels across different output channels $ch_{out}$. For such an architecture, the access rate is therefore mainly dependent on the input channel $ch_{in}$ and its firing activity. This simplifies the allocation to the non-uniform memory, as all kernel weights for a given input channel can be stored in the same region of the non-uniform weight memory. Knowing the size of each convolutional layer, the SNNs controller then only has to iterate over the input channels in order of descending fire activity, so that as the weight address is incremented it follows the expected average access rate to each address.

\subsection{Memory Control Logic}
\label{sec:memory control logic}
\begin{figure}[h]
\centering
\includegraphics[width=\columnwidth]{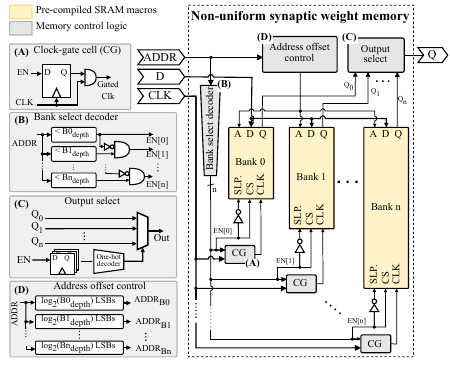}
\caption{Overview of the memory architecture for a configuration along with memory control logic.}
\label{fig:hw architecture}
\end{figure}

A circuit overview of the proposed non-uniform memory architecture is shown in Fig. \ref{fig:hw architecture}. The weight memory is organized into individually addressable, pre-compiled, SRAM macros of varying sizes (Bank 0 - Bank n). Control logic consists of clock gating cells (CG) (A), bank select decoders (B), output select logic (C), and address offset control (D).

The individual control logic blocks operate as follows: 
\begin{enumerate}[label=(\Alph*)]
    \item The clock-gating cell assures glitch-free disabling of the clock line to inactive memory banks. The gating of the individual memory clocks is controlled by the respective enable signal of that bank.  
    \item The bank select decoder outputs a one-hot enable signal to the corresponding memory bank based on the input weight address by comparing it to the depth of each memory bank. 
    \item The output select logic drives the output Q with the synaptic weight accessed from the previously selected memory bank. The selection is based on the enable signal from the previous access cycle, to account for the time required between applying an address at a bank's port and receiving the corresponding data at the banks output.
    \item The address offset control calculates the correctly offset address for each individual memory bank so that the concatenated memory bank forms a continuous address space from the view of the SNN controller.
\end{enumerate}

% Idk i guess no time for this section
%\subsection{Automatic Generation of Non-uniform memory bank}
%To extensively evaluate all possible non-uniform memory configurations a automated design flow that combining application requirements with hardware constraints is created. The memory generation flow is shown in Fig. X. Based on the target TOTAL synaptic memory size, the available 

\section{Results}
\label{sec:results}
\subsection{Benchmark Network Setup}
%To evaluate the non-uniform synaptic weight memory we train and deploy a set of spiking convolutional and fully-connected networks with LIF neurons, shown in table \ref{tab:networks}. To show the generalization of the proposed method across datasets, the networks target four datasets: image recognition (N-MNIST \cite{nmnist}), gesture recognition (DVSGesture \cite{DVSgesture}) and two KWS tasks (GSC \cite{GSC} and HD \cite{hd_speech}). The GSC and HD datasets were pre-processed with the speech2spikes algorithm \cite{speech2spikes} to give spiking input data. The DVSGesture and N-MNIST were cropped and dowmsampled to a resolution 32x32.

The evaluation of the non-uniform memory partition scheme is conducted by targeting four different datasets: image recognition (N-MNIST \cite{nmnist}), gesture recognition (DVSGesture \cite{DVSgesture}), and two KWS tasks (GSC \cite{GSC} and HD \cite{hd_speech}). Both GSC and HD datasets were pre-processed using the speech2spikes algorithm \cite{speech2spikes} for two fully connected SNNs, while DVSGesture and N-MNIST were cropped and downsampled at a resolution of 32x32 for two convolutional networks. The network topologies are described in table \ref{tab:networks}.

\begin{table}[h]
\centering
\caption{Networks used for benchmarking.}
\label{tab:networks}
\renewcommand{\arraystretch}{1.5} % Adjust the row height
\begin{tabular}{llll}
\hline
\textbf{Dataset (Net)} &
  \textbf{Architecture} &
  \textbf{Size} &
  \multicolumn{1}{c}{\textbf{Acc. (\%)}} \\ \hline
\textbf{N-MNIST (1)} &
  \begin{tabular}[c]{@{}l@{}}8conv5-MP2-14Conv5-\vspace{-4pt}\\ MP2-28Conv5-AP2-512FC11\end{tabular} &
  32KB &
  94.9\% \\
\textbf{DVSGesture (2)} &
  \begin{tabular}[c]{@{}l@{}}32conv5-AP2-58Conv5-\vspace{-4pt} \\ AP2-1450Fc11\end{tabular} &
  64kB &
  91.2\% \\
\textbf{GSC (3)} & 40Fc160-160Fc160-160Fc35 & 64KB & 83.5\% \\
\textbf{HD (4)}  & 40Fc120-120Fc120-120Fc20 & 32KB & 95.3\% \\ \hline
\end{tabular}
\end{table}

All networks were trained with snnTorch \cite{snntorch} using LIF neuron models and trained using backpropagation with a surrogate gradient descent. The weights were quantized to 8 bits. The two larger networks (2) \& (3) have a footprint of roughly 65000 parameters, allowing them to be deployed in a 64KB synaptic weight memory. The smaller networks (1) \& (4) are sized to fit in a 32KB synaptic weight memory. The same networks, with the same level of quantization, are deployed on both the traditional and the proposed non-uniform memory architectures, resulting in no difference in accuracy between the two.

\subsection{Memory Access Distribution}
\label{sec:memory access distribution}
After training each network, the firing activity of each neuron was recorded over the entire test datasets. The memory access rate of each synaptic weight was derived from the recorded firing activities and the memory access order described in \ref{sec:address allocation}. The subset of accesses to the most frequently read weights, as a share of all weights in the network, is shown in table \ref{tab:access distribution}. 

\begin{table}[h]
\centering
\caption{Share of total weight accesses to subset of most frequently accessed weights.}
\label{tab:access distribution}
\renewcommand{\arraystretch}{1.5} % Adjust the row height
\begin{tabular}{lccc}
\hline
\multirow{2}{*}{\textbf{Dataset (Net)}} & \multicolumn{3}{c}{\textbf{\begin{tabular}[c]{@{}c@{}}Share of all memory accesses from \vspace{-4pt} Y\% most\\ frequently accessed weights\end{tabular}}} \\
 & \textbf{Y = 10\%} & \textbf{Y = 25\%} & \textbf{Y = 50\%} \\ \hline
\textbf{N-MNIST (1)}    & 54.1\% & 63.2\% & 77.2\% \\
\textbf{DVSGesture (2)} & 71.0\% & 79.4\% & 87.5\% \\
\textbf{GSC (3)}        & 33.8\% & 62.4\% & 88.7\% \\
\textbf{HD (4)}         & 35.1\% & 73.8\% & 93.8\% \\ \hline
\end{tabular}
\end{table}

The results show a large variety in the access rate of synaptic weights across both fully connected and convolutional networks. For the networks trained on KWS tasks, roughly a third of all memory accesses are to only 10\% of the most frequently accessed weights. The convolutional networks show even more imbalance in the access rate, with 54.1\% and 71.0\% of accesses being at the top 10\% of the weights for the N-MNIST and DVS networks, respectively. 

%The observed non-uniformity in memory accesses rates is likely caused by the general fact that sparsity increases with the depth of the network, leading to weights in later layers naturally being less frequently accessed (assuming sparsity is utilized efficiently) \cite{imec_review}. %Furthermore, for convolutional networks, spatial feature size tends to decrease while channel count tends to increase with the depth of the network.

\subsection{Design Space Exploration}
An automated design space exploration was set up to evaluate non-uniform synapse memories and find the optimal configuration based on application and hardware constraints. For all networks evaluated, we assumed that a single 8-bit synaptic weight is stored per addressable word in the memory. This was based on the fact that to leverage the high sparsity of spikes, an efficient SNN should be able to load single synaptic weights at a time. Otherwise, upon receiving a single spike, it may have to load weights corresponding to pre-synaptic neurons which have not fired. The used SRAM macros were limited to sizes between 1 KB and 32 KB with the upper limit being set by the provided memory compiler and the lower limit by the observation that there was no further reduction in access power by reducing the memory size further. %Furthermore, the size of the used SRAM macros was constrained to depths that are powers of two to fully utilize the periphery logic within each macro. 

For the smaller 32 KB synapse memory, a total of 202 possible ways to assemble a concatenated 32 KB memory using the available smaller macros were found. For the 64 KB synapse memory, the possible combinations had to be constrained to be reasonably evaluated, resulting in 686 combinations. For each of the found combinations, a synthesizable Verilog code was automatically generated. The code instantiates the available memory macros along with the memory control logic described in section \ref{sec:memory control logic}.

% Needs to be updated with new training data!!
\subsection{Synthesis Setup}
The generated Verilog code from the design space exploration for both the uniform and non-uniform memories was synthesized in 28-nm CMOS technology. The individual memory macros were generated using the foundry-provided memory compiler tools. All evaluated designs were synthesized at 100 MHz using libraries in the Typical-Typical (TT) corner at 25$^{\circ}$C and a 1 V supply voltage. To accurately evaluate the read access power of each memory, a switching activity file was generated for each benchmark network to model the SNN controller and its read requests to the synaptic weight memory. The access pattern was based on the extracted average neural firing activities as described in \ref{sec:memory access distribution} and uses the address allocation and access scheme described in \ref{sec:address allocation}. 

\subsection{Access Power Reduction and Area Overhead} The quantified values of both area and access power for select well-performing non-uniform memories from the design space exploration are shown in table \ref{tab:syn_result_table}. Both access power reduction and memory area overhead are represented in percentage change over a conventional synapse memory consisting of a single SRAM bank.  %For uniform weight memories, the average power is identical for both evaluated datasets as the access cost to the memory is the same regardless of the address. 

\begin{table}[h]
\centering
\caption{Synthesis results for select memory configurations from the design space exploration. Area and power is normalized and in relation to the size and power of a conventional synapse memory consisting of a single SRAM macro.}
\label{tab:syn_result_table}
\renewcommand{\arraystretch}{1.5} % Adjust the row height
\begin{tabular}{p{.3cm} p{.6cm} p{1.4cm} p{2.3cm} p{1.9cm}}
\hline
\textbf{ID} & \textbf{Size} & \textbf{Content} & \textbf{Power Reduction} & \textbf{Area Overhead} \\ \hline
\multicolumn{5}{c}{Proposed Non-Uniform Memories} \\ \hline
A & 64KB & 1$\times$32KB  1$\times$16KB  1$\times$8KB  2$\times$4KB & DVS: 49\% \par GSC: 40\%    & 8\% \\
B & 64KB & 2$\times$16KB  2$\times$8KB  7$\times$2KB  2$\times$1KB  & DVS: 57\% \par GSC: 49\%    & 16\% \\
C & 32KB & 1$\times$16KB  2$\times$8KB                              & N-MNIST: 39\% \par HD: 38\% & 13\% \\
D & 32KB & 2$\times$8KB  8$\times$2KB                               & N-MNIST: 63\% \par HD: 61\% & 18\% \\ \hline
\multicolumn{5}{c}{Uniform Memories} \\ \hline
E & 64KB & 4$\times$16KB                                            & 23\%                     & 4\% \\
F & 64KB & 8$\times$8KB                                             & 38\%                     & 23\% \\
G & 32KB & 4$\times$8KB                                             & 37\%                     & 22\% \\
H & 32KB & 8$\times$4KB                                             & 58\%                     & 38\% \\

\hline
\end{tabular}
\end{table}

The non-uniform configurations A \& B form a 64KB memory and are evaluated on network 2 \& 3 in table \ref{tab:networks}. Configurations C \& D form a 32KB memory and are evaluated on networks 1 \& 4. Reported read access power includes both dynamic and leakage power during inference.

% \begin{figure}[h]
% \centering
% \includegraphics[width=\columnwidth]{figures/results_single_axis_barplot.pdf} 
% \caption{Synthesis results for select memory configurations from the design space exploration. Area and power are normalized to the size and power of a single uniform memory macro. The select non-uniform memories A-D consist of the following memory bank sizes. 
% A: 1$\times$32KB + 1$\times$16KB + 1$\times$8KB + 2$\times$4KB 
% B: 2$\times$16KB + 2$\times$8KB + 7$\times$2KB + 2$\times$1KB, 
% C: 1$\times$16KB + 2$\times$8KB 
% D: 2$\times$8KB + 8$\times$2KB}
% \label{fig:power results}
% \end{figure}

\section{Discussion}
\label{sec:discussion}
\textbf{Power/Area trade-off:} Compared to a single 32KB-sized synaptic weight memory, the non-uniform architectures are able to reduce memory access power by upwards of 63\%, at an 18\% area overhead. While segmenting the weight memory into uniformly sized banks can achieve similar results in terms of access power reduction, this tends to come at a much greater area overhead, as some of the banks go heavily underutilized. The 32KB configuration D for example, is able to achieve a greater access power reduction than the uniform 8x4KB memory, at a 2.1$\times$ lower area overhead. 

\textbf{Control logic overhead:} Across all of the select non-uniform memory configurations, the area of the control logic made up less than one percentage of the total memory area. The control logic requires no additional clock cycles of delay for a memory access, as the added delay through the added control logic is negligible compared to timing constraints of the memory macros themselves, making the proposed design suitable also for real-time applications.

\textbf{Variation across network architecture \& datasets:} As the share of memory accesses to the most frequently accessed weights in the convolutional networks targeting N-MNIST and DVS is larger than in the fully connected networks, these networks perform better than the fully connected ones on the non-uniform memories in the select benchmarks. At most, the difference in access power reduction between any of the benchmarks is around 14\%, showing that the proposed architecture can generalize well across varying network sizes and topologies.

\textbf{Flexibility:} A non-uniform memory architecture can provide greater flexibility in deploying networks of varying sizes, as unlike traditional synapse memories, which require full memory occupancy for optimal use, smaller networks can be efficiently deployed in more shallow memory regions, with unused macros disabled for improved efficiency.

\section{Conclusion}
In this work, we presented how a non-uniform memory architecture can be applied to SNNs to reduce the memory access cost by allocating synaptic weight to memory banks of varying sizes based on average neural firing activities. A detailed circuit-level architecture is proposed along with an efficient address allocation scheme to eliminate the need for bank lookup logic. Results across multiple networks and datasets have demonstrated that our proposed architecture can reduce memory access power drastically compared to a single synaptic weight memory bank, at a significantly lower area overhead as compared to a uniformly partitioned memory.

\section*{Acknowledgement}
This work was supported by Innovationsfonden through the CORROSENSE Project under Grant 2081-00027A.

\newpage
\bibliographystyle{IEEEtran}
\bibliography{references}

\end{document}